# Comparison of compact bone failure under two different loadings rates: experimental and modelling approaches.

M. Pithioux[⊗] & D. Subit[*+], P. Chabrand[*⊗].

[*] Laboratoire de Mécanique et d'Acoustique, CNRS, 31 Ch. Joseph Aiguier 13402 Marseille Cedex 20, France

[+] INRETS, Laboratoire de Biomécanique Appliquée, bd. P. Dramard, 13916 Marseille, Cedex 20, France and GDR 2610 "Biomécanique du choc"

[⊗] Laboratoire d'Aérodynamique et de Biomécanique du Mouvement, CNRS, 163, avenue de Luminy - Case 918 -13288 Marseille Cedex 9, France and GDR 2610 "Biomécanique du choc"

mail: {pithioux,subit,chabrand}@morille.univ-mrs.fr

tel: +33 (0)4 91 26 61 77

fax: +33 (0)4 91 41 16 91

# Comparison of compact bone failure under two different loadings rates: experimental and modelling approaches.

M. Pithioux[⊗] & D. Subit*[+], P. Chabrand*[⊗].

* Laboratoire de Mécanique et d'Acoustique, CNRS, 31 Ch. Joseph Aiguier 13402 Marseille Cedex 20, France

[+] INRETS, Laboratoire de Biomécanique Appliquée, bd. P. Dramard, 13916 Marseille, Cedex 20, France and GDR 2610 "Biomécanique du choc"

[⊗] Laboratoire d'Aérodynamique et de Biomécanique du Mouvement, CNRS, 163, avenue de Luminy - Case 918 -13288 Marseille Cedex 9, France and GDR 2610 "Biomécanique du choc"

mail: {pithioux,subit,chabrand}@morille.univ-mrs.fr

**Abstract:**

Understanding the mechanical behaviour of bones up to failure is necesary for diagnosis and prevention of accident and trauma. As far as we know, no authors have yet studied the tensile behaviour of compact bone including failure under dynamic loadings (1m/s). The originality of this study comes from not only the analysis of compact bone failure under dynamic loadings, the results of which are compared to those obtained under quasi static loadings but also the development of a statistical model. We developed a protocol using three different devices. Firstly, an X-ray scanner to analyse bone density, secondly, a common tensile device to perform quasi static experiments and thirdly, a special device based upon a hydraulic cylinder to perform dynamic tests. For all the tests, we used the same sample shape which took into account the brittleness of the compact bone. We first performed relaxation and hysteresis tests followed by tensile tests up to failure. Viscous and plastic effects were not relevant to the compact bone behaviour so its behaviour was considered elastic and brittle. The bovine compact bone was three to four times more brittle under a dynamic load than under a quasi static one. Numerically, a statistical model, based upon the Weibull theory is used to predict the failure stress in compact bone.

## 1. Introduction

Bone failure often occurs in accidental shocks such as locomotion, races, sports or coach accidents. Cortical bone shows tearing, damage and failure mechanisms when it receives a shock. In the bibliography [1-10], damage and failure of bovine compact bone have already been studied in quasi static tensile experiments and have shown large variations of stress failure, from 100 to 200 MPa, and failure strain, from 0.4 to 4 %. However, none of these authors had studied compact bone failure under dynamic loads. Other authors have performed studies in dynamic [11-12]. Various methods have been used (Hopkinson bar stress method and in vivo strain measurements) to characterise the human femoral cortical bone behaviour but for a non-damaging range of loadings. Only [13] tested compact bone sample under impact, for a velocity of 0.3 m/s. This velocity is not high enough for application in accidentology and traumatology.

The aim of this paper is to compare compact bone failure under quasi static and dynamic tensile loads. A statistical law was used to analyse the stress failure variation and was useful in the development of a constitutive law for compact bone. With this aim, we created a complete protocol comprising the following steps: firstly, the bone structure was analysed using an X-ray scanner to determine where the structure was homogeneous, secondly, hysteresis and relaxation tests were performed to qualify the mechanical behaviour of the bone, and thirdly, tensile tests up to failure were carried out; namely, quasi static experiments with a speed range of 0.5 mm/min, 5 mm/min, 10 mm/min and 500 mm/min and dynamic experiments with a speed of 1m/s. For all the experiments, tensile strength and displacement were measured. Finally, we analysed the failure using a statistical model that took into consideration the biological variability of compact bone behaviour.

## 2. Material and method

### 2.1. Determination of sample structures

We analysed bone structure by using an X-ray scanner, type ND8000, Laboratoire de Mécanique et d'Acoustique. The samples were taken from 20 fresh bovine femoral bones. Animals were from five to seven years old at the time of death. The bones were frozen prior to the experiments. The epiphyses were cut off so that we could concentrate our attention only on compact bone. The diaphyses were about 110 mm long and using the X-ray scanner, we cut 1 mm thick slices every 10 mm (Fig.1). The scanner was then calibrated to estimate the radiological density of bone (CT unit). We found two types of sections. Type I sections where the radiological density was 900 CT (± 10 CT), and type II sections where the radiological density varied from 800 CT to 1100 CT. We then used an optical microscope to analyse the section structure more accurately. Type I samples had a lamellar structure, whereas type II samples had an osteonal structure. We chose to work on type I samples that were as homogeneous as possible in order to reduce the number of relevant parameters which explained failure process. We assumed that density had a large influence on failure process, so we worked on samples with a low density variation assuming, moreover, that the results depended mainly on the presence of defects. These defects may have been characterised by a variation in the porosity or in the mechanical characteristics of bones.

## 2.2. Sample shape

Samples were cut in areas where the bone was homogeneous (type I), namely in the anterior lateral and anterior medial parts. Bone shafts were taken and cut in the axial direction and marrow was removed from each part. Samples were then machined with a numerically controlled machine tool. As is classically done for tensile samples, the sample width was reduced around the centre to localise failure in this part. However, in this case the width was gradually reduced and shaped as shown in figure 2 to avoid the appearance of failure close to the extremities. After this process, the samples are dry.

## 2.3. Qualitative study

We first performed tests to qualify the model of mechanical behaviour that could describe the compact bone behaviour. There are two kinds of loading, namely hysteresis and relaxation.

### 2.3.1 Hysteresis experiments

In this part, we studied plasticity of bone. Tensile experiments were carried out using a common tensile device -Instron, Fig. 3(a). We imposed cycles every 0.1% of strain up to 0.5%. The realised displacements and the forces were measured. The displacement of the lower traverse beam was measured using an LVDT sensor (Linear Variable Differential Transformer), attached to the machine frame. The tensile load was measured by a strain gauge sensor with an uncertainty of measurement of 1% on the upper traverse beam. These tests were performed on 3 samples.

### 2.3.2. Relaxation tests

Relaxation tests are a good way of observing viscous behaviour. In these experiments, displacement, equivalent to strain, was prescribed and the force variations, equivalent to stress variations, were measured. Six samples were tested. We prescribed a strain of 0.5 % (0.3 mm), and a strain rate of 0.26 %/second. The stress was then measured during 60 seconds.

2.4. Quantitative tests

Tensile tests were performed up to failure to identify the value of the mechanical quantities (Young's modulus E and failure stress $\sigma_{ult}$) used to describe the behaviour of the compact bone.

2.4.1. Quasi static experiments

The experiments were carried out on the same tensile device we used for the hysteresis tests, (Fig. 3(a)), with imposed displacement (speeds of 0.5 mm/min, 5 mm/min, 10 mm/min and 500 mm/min). The sampling rate of the data acquisition was 10 Hz with tests lasting for more than 7 seconds.

2.2 Dynamic experiments

We designed a special device to perform dynamic tensile tests (velocity: 1m/s). It was attached to a hydraulic jack fixed onto the upper traverse beam, Fig. 3(b). The force sensor was placed under the lower chuck jaw and experiments were carried out using an imposed displacement**.** The realised displacements, velocities, and accelerations were measured, with upper chuck jaw displacements and velocities being measured using a laser vibrometer. Displacement was obtained by interference measurements with an uncertainty of measurement of 1 %, whereas velocity was measured by the Doppler effect and acceleration by an accelerometer fixed onto the upper chuck jaw. Force was measured by a triaxial piezoelectric sensor set on the lower chuck jaw, which measured forces in the tensile direction, with an uncertainty of measurement of 4.7 %, and in the shear plane. The hydraulic device was validated by carrying out tests on known materials. Measurements in the shear plane showed *a posteriori* that tensile force was much greater than flexion and shearing forces. An initial displacement of 1/100 mm, that is to say a 0.016 % global strain, was prescribed in order to avoid dynamic effects due to the assembling of the system, especially the clearance. The sampling rate of the data acquisition was 32 kHz with tests lasting about 3 ms. This high sampling rate led us to use the laser vibrometer whose cut-off frequency was higher than the sampling rate.

## 2.5. Statistical model

A statistical model based on the Weibull theory was developed to analyse tensile results [14; 15]. Let V be the reference volume, being constant for all the samples; let $P_R(\sigma)$ be the failure probability of the volume V subjected to one-dimensional tensile stress σ, that is to say the failure probability of the sample. Using the function *f*, the failure probability was given by:

$$P_R(\sigma) = 1 - exp(-f(\sigma)) \tag{1}$$

To give an approximation of this unknown function *f*, Weibull proposed the following function [16-20]:

$$f(\sigma) = \left(\frac{\sigma}{\sigma_0}\right)^m \tag{2}$$

$\sigma_0$ is the Weibull statistical failure stress for the considered sample set.

Using the natural logarithm in equation (1), according to equation (2), we have:

$$ln\left[ln\left(\frac{1}{1-P_R(\sigma)}\right)\right] = m\, ln\,\sigma + K \tag{3}$$

$$with\ K = m\, ln(\sigma_0) \tag{4}$$

Equation (3) represents a straight line whose slope is *m*. *m* was named Weibull's modulus. As presented above, the stresses, namely the ratio of force to section where failure occurred, were plotted against the strains, namely the ratio of displacement to initial length. Data obtained experimentally showed the behaviour up to failure. The statistical model was applied to the ultimate stress ($\sigma_0 = \sigma_{ult}$) to obtain the probability law of failure stress.

# 3. Results

## 3.1. Qualitative tests

### 3.1.1. Hysteresis tests

From these tests we concluded that plasticity could be disregarded when describing bone behaviour. At the end of the cycle, there was no residual plastic strain, the stress-strain curve (Fig 4) was linear. In addition, the facies failure (Fig 5) confirmed that compact bone is a brittle material as no plastic strain was observed.

### 3.1.2. Relaxation tests

Results are presented in figure 6. They show that only an elastic return was observed after the sample was placed under tension. Relaxation effects were not significant enough, allowing us to conclude that compact bone material does not relax.

In conclusion, the qualitative tests justified the use of an elastic model for the compact bone in quasi-static and dynamic cases.

## 3.2. Quantitative tests

### 3.2.1. Tensile experiments under quasi static loads

From these results, the mechanical properties of compact bone were deduced where failure occurred. Stress-strain curves were divided into three parts (Fig 7). On the first part of the curve where the behaviour was linear elastic, Young's modulus could be calculated. On the second part, one could observe that the behaviour became weakly non-linear, showing that the material was damaged. In the final stage failure occurred suddenly. The damaging part of the bone behaviour could be neglected as this behaviour was not relevant to describe bone failure, because the non linearity is weak. Hence, this behaviour was represented by a brittle elastic model, and was defined by Young's modulus E and the ultimate stress $\sigma_{ult}$.

Results are presented in table 1. Failure strain and stress varied greatly from one quasi-static experiment to another, so the failure stress values probably depended on the distribution of defects in the sample. The largest defect in the structure could have been the one that caused the failure. The measured failure stresses varied by 84%, and Young's Modulus by 55%. These results however are in line with those published [1; 4; 5; 7; 21; 22]. This broad variation justified the use of a statistical model that takes into account the whole range of behaviours that we observed.

For each experimental result $\sigma_{ult}$, we plotted $ln\left[ln\left(\frac{1}{1-P_R(\sigma_{ult})}\right)\right]$ versus $ln(\sigma_{ult})$. We determined Weibull's coefficients *m* and *K* by a least square linear regression. We found *m*=5.77 and *K*=-29.4, $R^2$ = 0.94 where $R^2$ was the estimator of the least square method.

The failure probability law was then defined by:

$$P_R(\sigma_{ult}) = 1 - e^{-\left[\frac{\sigma_{ult}}{163.3}\right]^{5.77}} \quad (5)$$

This law is plotted in figure 8(a). The failure probability defines a constitutive law of compact bone.

Using this law, we gained statistical information on failure probability. For example, we found that there was no possibility of sample failure for a stress lower than 50 MPa, a 46% chance of sample failure for a stress lower than 150 MPa and a sample failure for stresses greater than 220 MPa was certain in our study case. In fact, the number of samples is weak and results depend on the biological variability of bone behaviour.

### 3.2.2. Tensile experiments under dynamic loads

When experimenting, we found the same three parts as in the quasi static case for bone behaviour (Fig 9) and used a brittle elastic behaviour model. As in quasi-static, no plastic strain was observed on the failure facies. Only the second part, where bone behaviour was non-linear and where the bone was damaged, was more obvious. As we were interested in bone failure, we used the same elastic and brittle model of behaviour as in the quasi static case.

Results are summarized in table 1. As in quasi-static, the distribution of defects in the sample led to strain and stress failures which varied greatly from one experiment to another in dynamic. The largest defect in the structure may have been the one that caused the failure. Failure stresses varied by about 72%, which corresponded to a large variation under dynamic loads, and Young's Modulus by about 115%. A statistical model was then useful as in quasi static cases. Stress and strain failures were lower in the dynamic case than in the quasi static case. However, the same range of variations of Young's modulus values were obtained in the two cases.

We then identified the parameters of the statistical law as: $m$=7.3 and $K$=-27.1, $R^2$ = 0.98. The failure probability law was thus defined by:

$$P_R(\sigma_{ult}) = 1 - e^{-\left[\frac{\sigma_{ult}}{41}\right]^{7.31}} \qquad (6)$$

The failure probability and the constitutive law of compact bone under dynamic load are plotted in figure 8(b).

We found, in this case, that there was no possibility of failure for a stress lower than 15 MPa, a 57 % chance of sample failure for a stress lower than 40 MPa, and a sample failure for a stress greater than 55 MPa was certain in our study.

**4. Conclusions and discussion.**

The originality of our approach consisted firstly, in studying bone failure under dynamic loadings (1m/s) secondly, in comparing these results with those obtained under quasi static loadings and thirdly, in developing a statistical model. Moreover this statistical model has been developed to be implemented in finite elements software to predict bone lesions due to an impact. To do so, we developed a protocol using three different devices by combining the use of a microscope and an X-ray scanner as well as quasi static and dynamic tensile devices. The same variation of results was obtained using the two devices with quasi static loads. The large range of variation observed in both cases and the brittle properties of compact bones led to the development of a statistical model. Failure stress in the two cases showed that the compact bovine bone was three to four times more brittle under dynamic load than under a quasi static one (see table 1). As the viscous effect was disregarded, the observed differences between quasi-static and dynamic behaviour were probably due to the fibrous bone structure. The compact bone fibres could have been responsible for the observed stress diminution. Indeed, in all the specimens, there were three types of lamellar structure whose fibres were oriented longitudinally, obliquely and transversally; that is to say, at 0°, 45° and 90° with respect to the longitudinal axis of the structure [23-26]. In the dynamic case, the velocity experiment may have prevented fibres from turning in the tensile direction and causing shear stresses along the fibres. Consequently fibres may have been damaged and become more brittle.

The same protocol could have been applied to healthy human bones or to pathological ones, such as, for example, osteoporotic or cancerous bones for example. Obviously this study is limited to the fact that *in vivo* the bone is embedded in a complex muscular and ligamentous system being totally different from the experimental setup in the study. However, our topic was a classical mechanical approach to identify bone characteristics in order to define bone behaviour and obtain data to insert them into a human virtual model. Dynamic loading are found in areas of human motion or accidentology where our work can be used. Examples are locomotion or racing and also in sports or car accidents.

This work, and particularly the extension of the statistical method, provides the possibility of predicting bone failure and could lead to applications in computer-assisted surgery (CAS). Only a small number of samples were tested, but it was sufficient enough to provide an estimation of the failure law. In parallel to these experiments, we developed a numerical model of the microscopic behaviour of compact bone at failure under quasi static and dynamic loads [14]. Our aim is now to apply the statistical model, obtained at the macroscopic level, to a microscopic model to predict the lesion occurrence and the damage propagation in the structure.

List of captions

Figure 1: X-Ray scanner image, (a) in axial direction and (b) in the orthogonal plane to the axial direction (scales were not observed). (a) white lines show the position where pictures such as (b) were taken. (b) white rectangles show the places where samples were cut.

Figure 2: Sample geometry.

Figure 3: (a) Quasi static device and (b) hydraulic device.

Figure 4: Hysteresis at (a) 0.1% and (b) 0.5% of strain.

Figure 5: Failure of a facies sample after dynamic testing (1m/s).

Figure 6: Compact bone relaxation according to time.

Figure 7: Quasi static stress-strain curves (ten samples).

Figure 8: Failure probability laws of compact bones under (a) quasi static and (b) dynamic loads.

Figure 9: Dynamic stress-strain curves (seven samples).

Table 1: Young's modulus (E), failure stress ($\sigma_{ult}$), and strain ($\varepsilon_{ult}$) variations obtained using quasi static and dynamic devices.

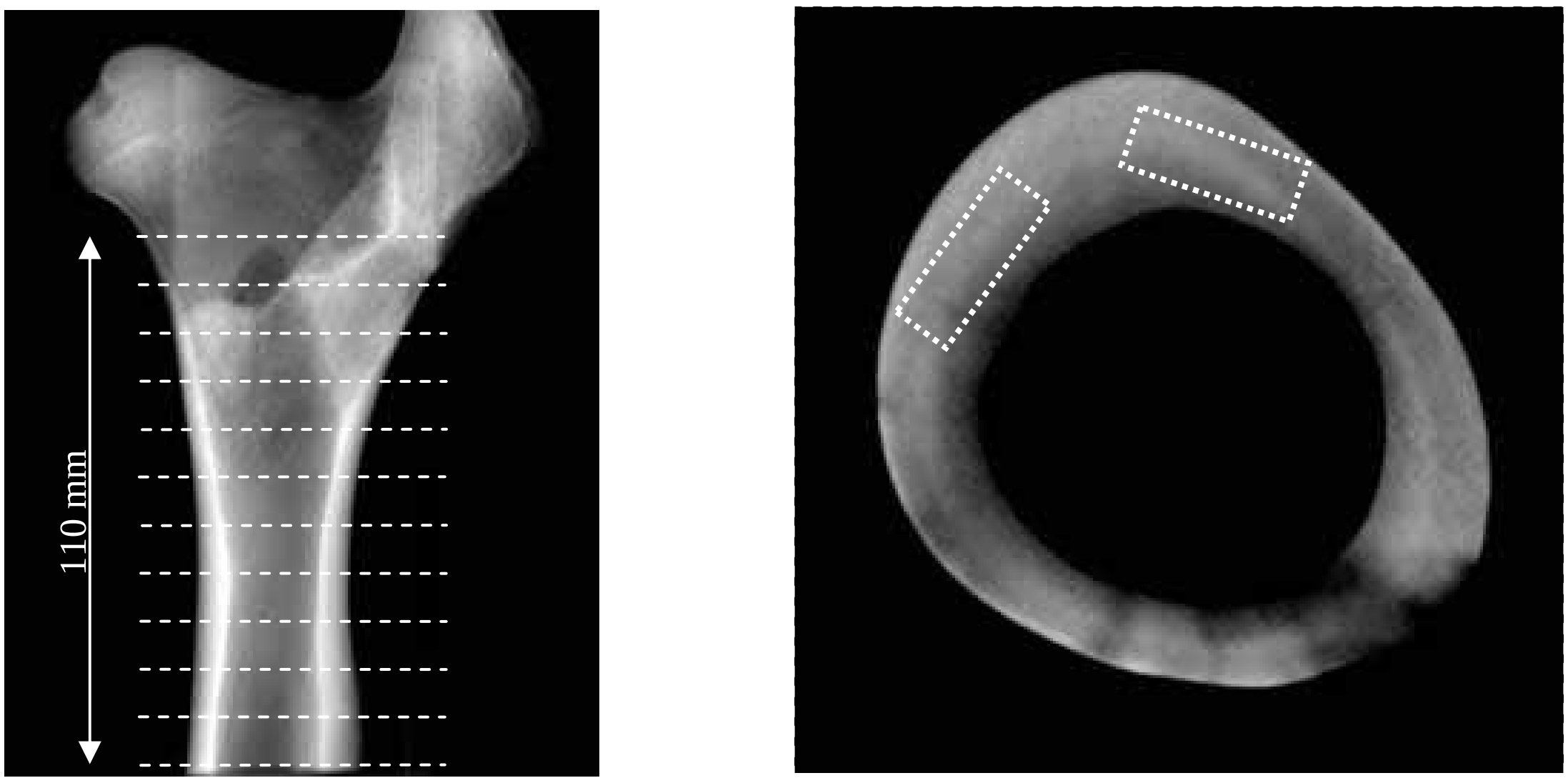

(a) (b)

Figure 1

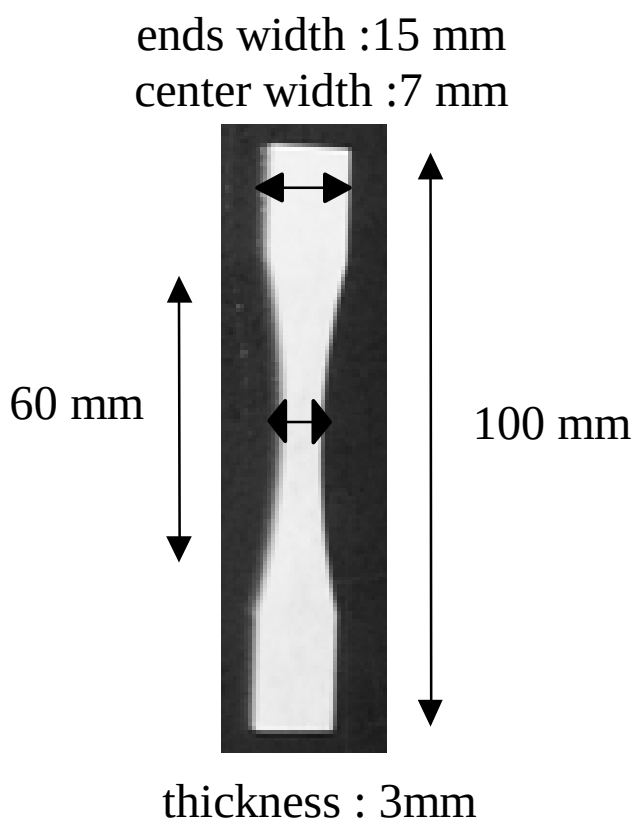

Figure 2

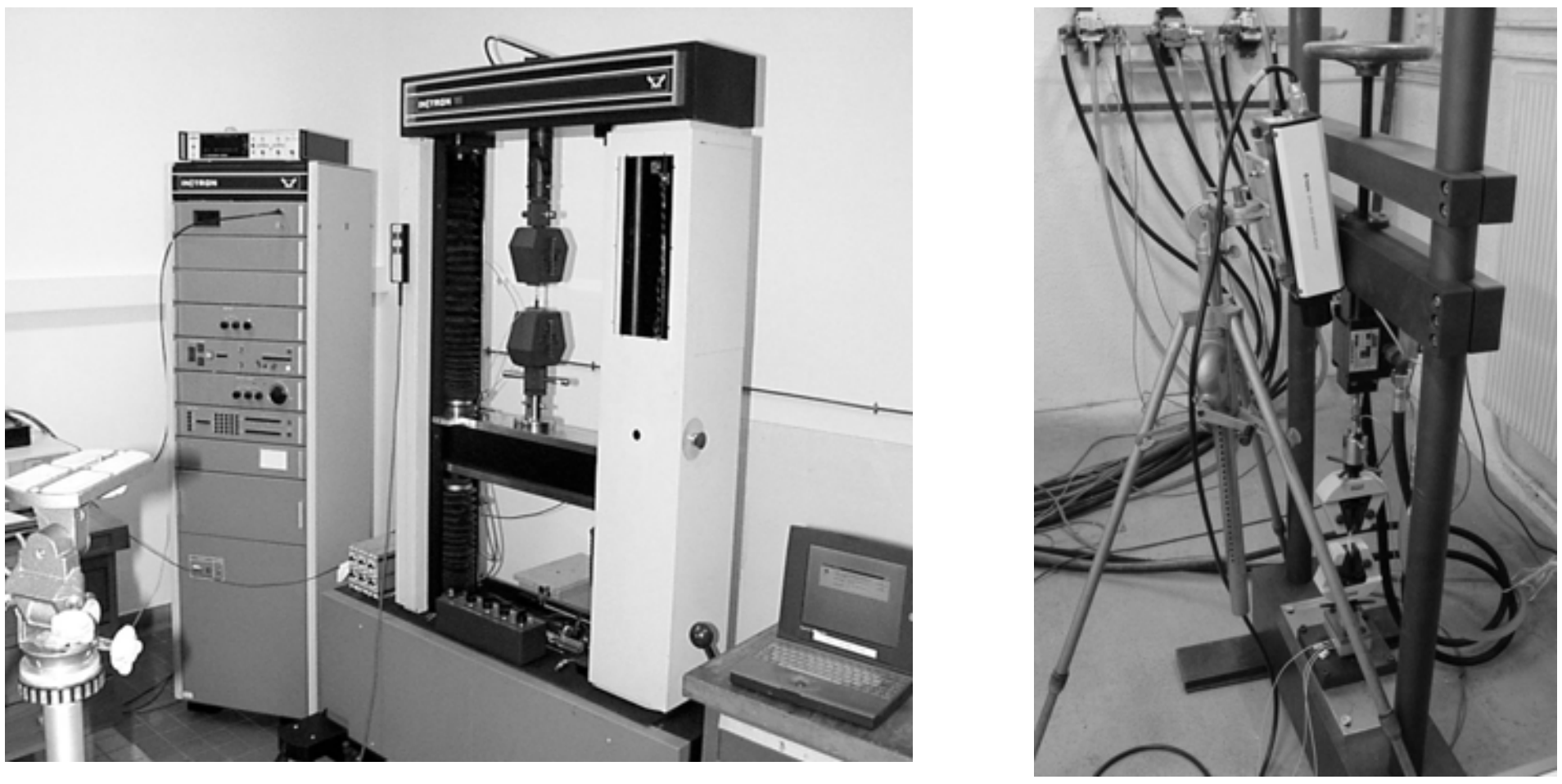

(a) (b)

Figure 3

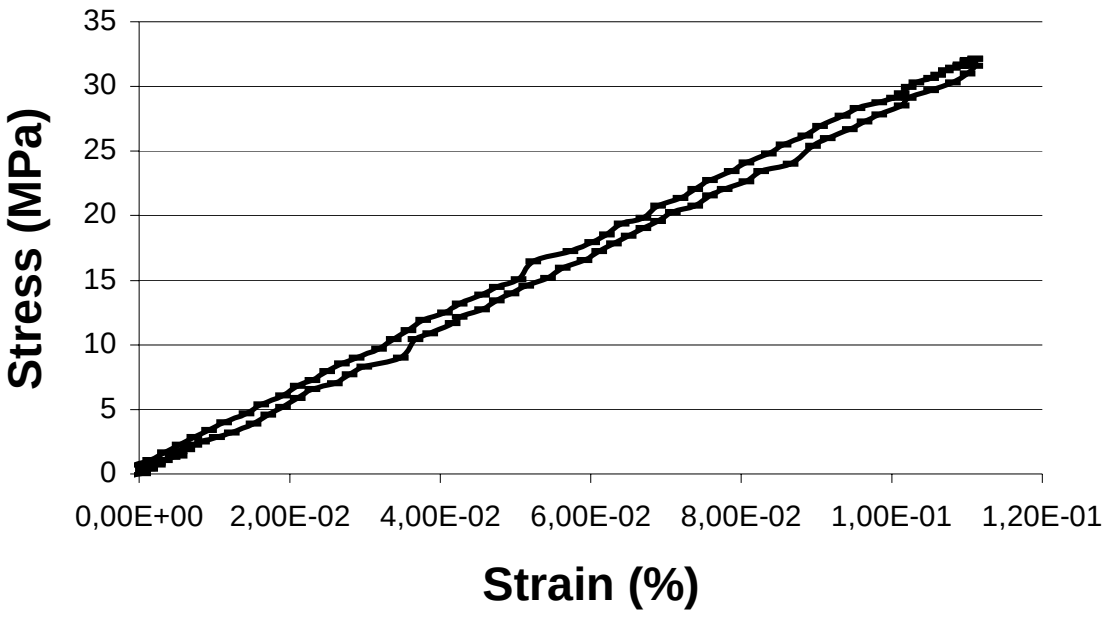

(a)

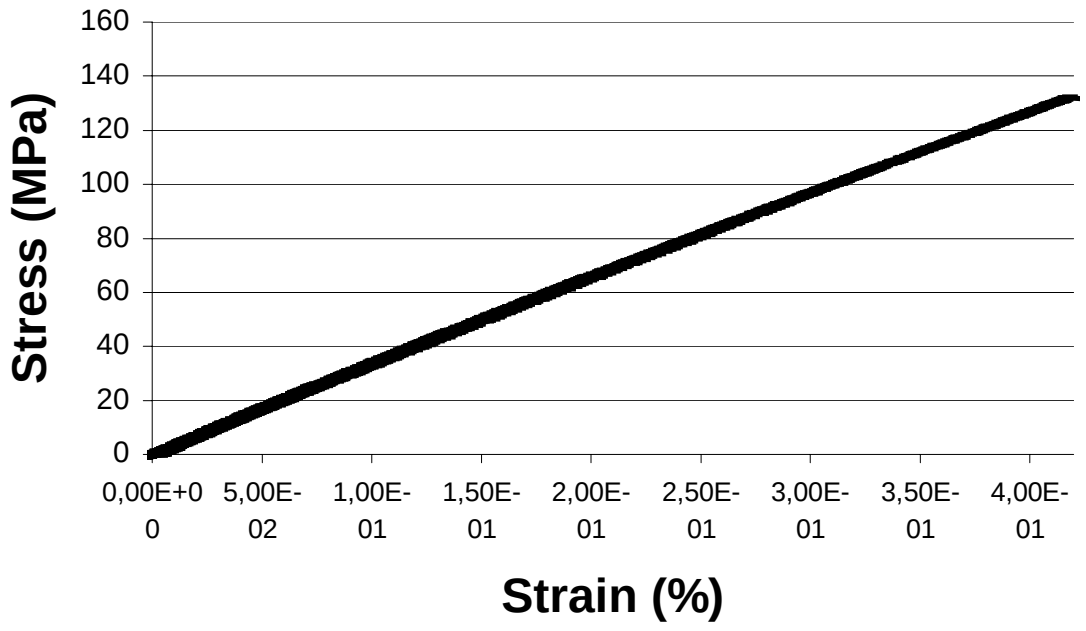

(b)

Figure 4

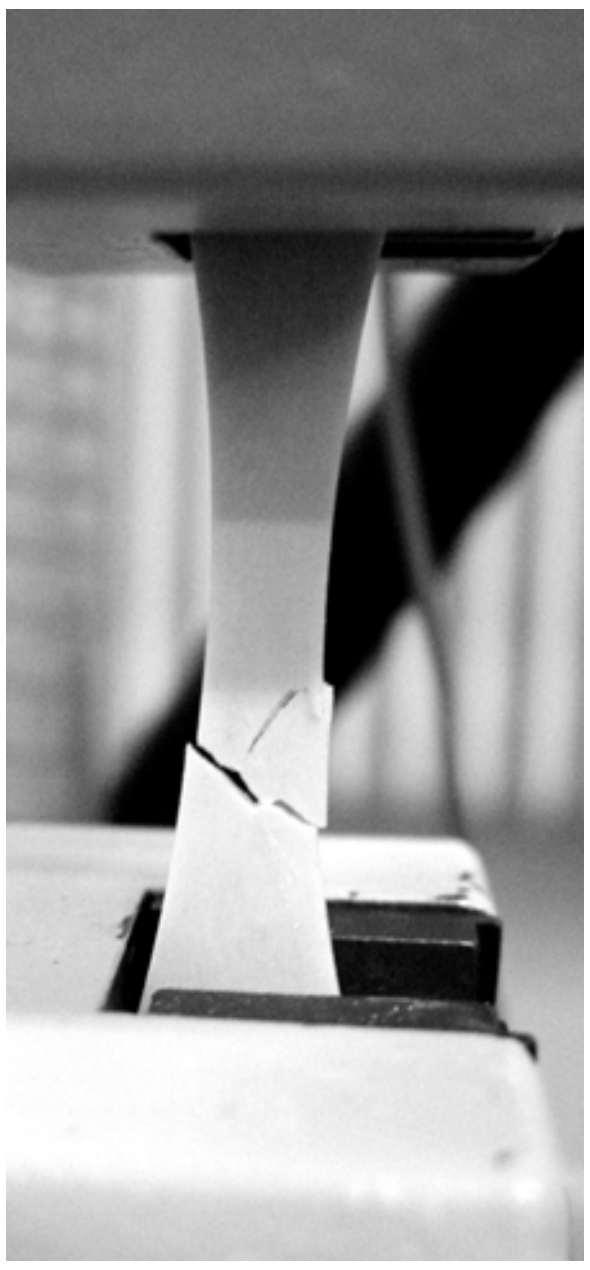

Figure 5

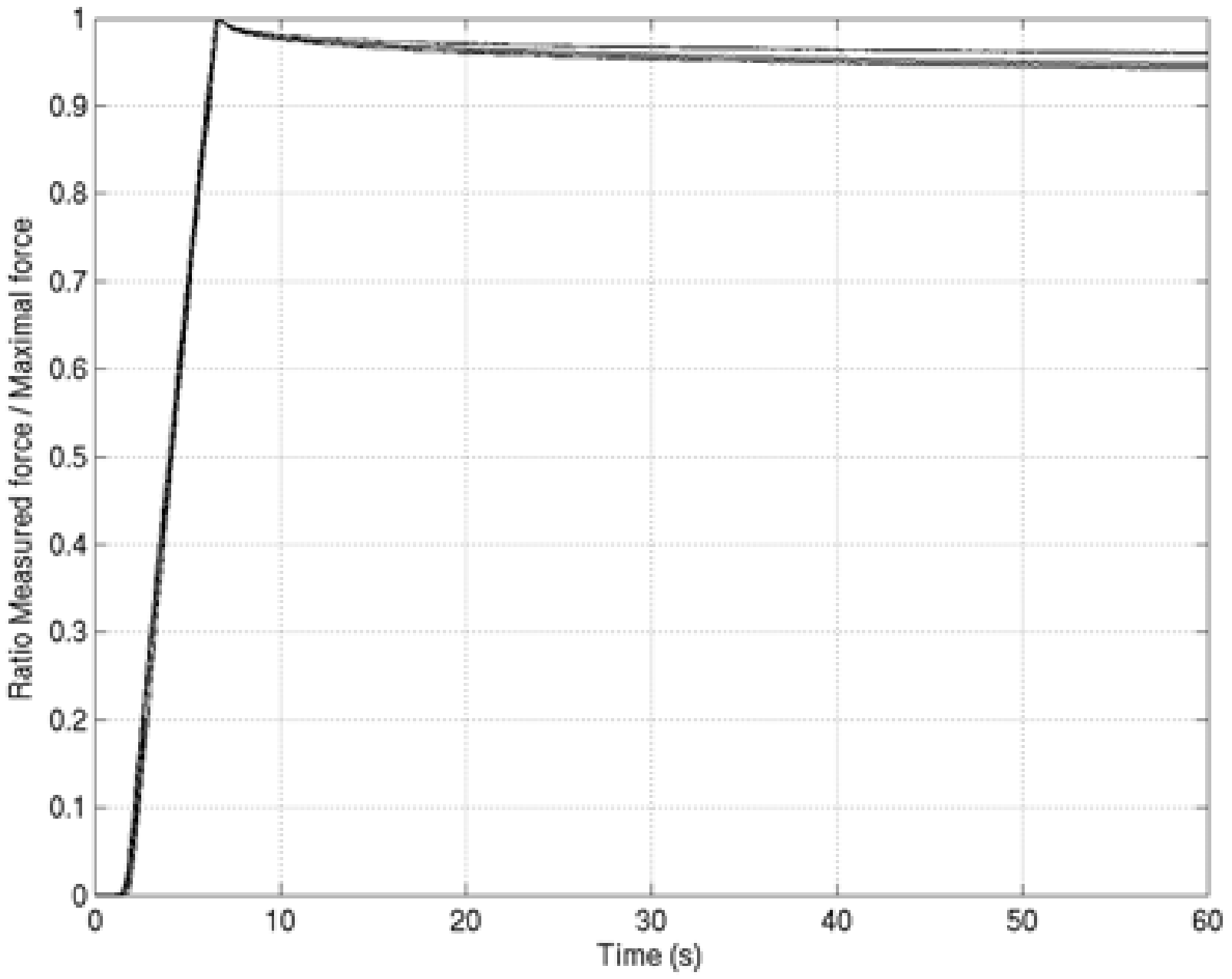

Figure 6

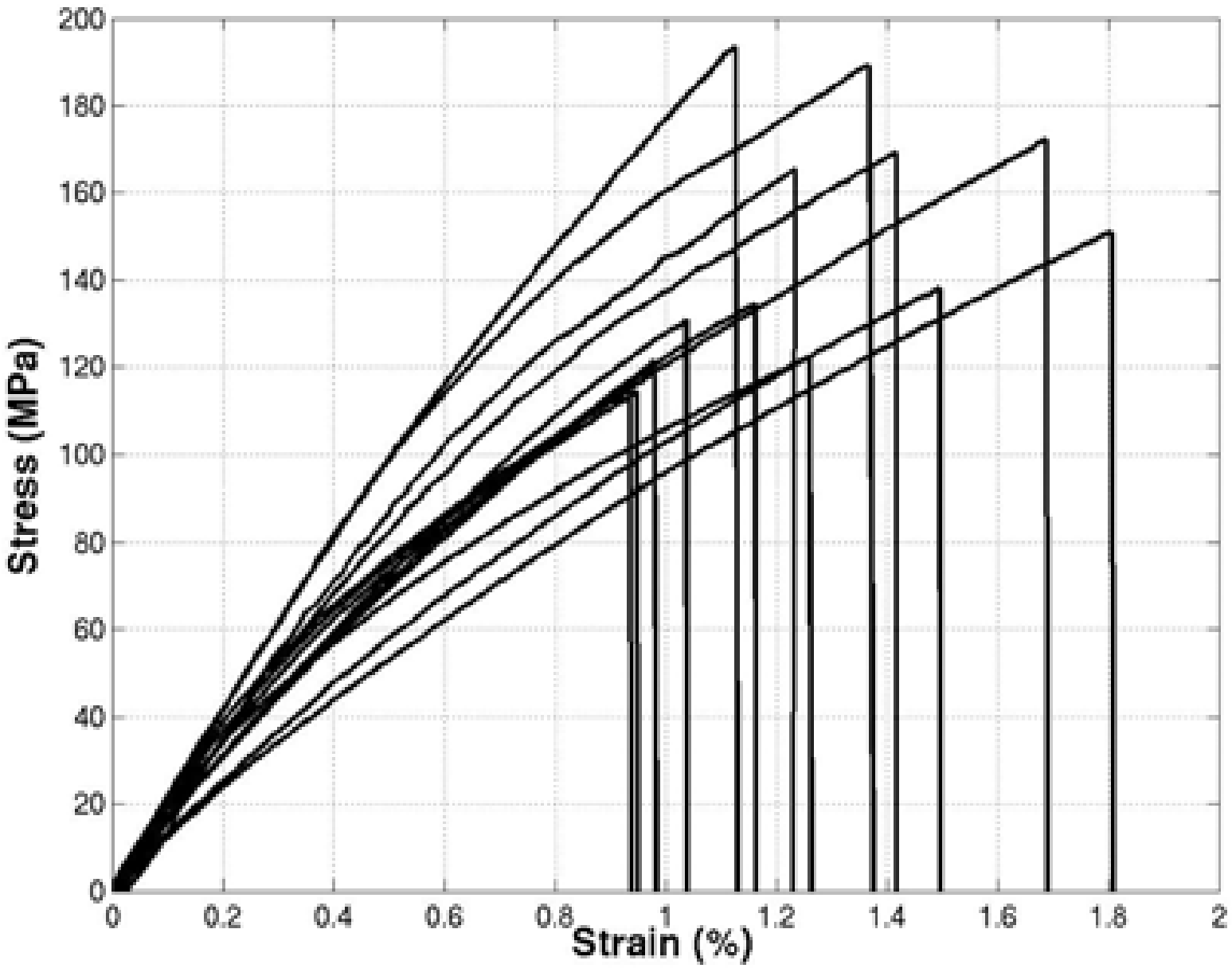

Figure 7

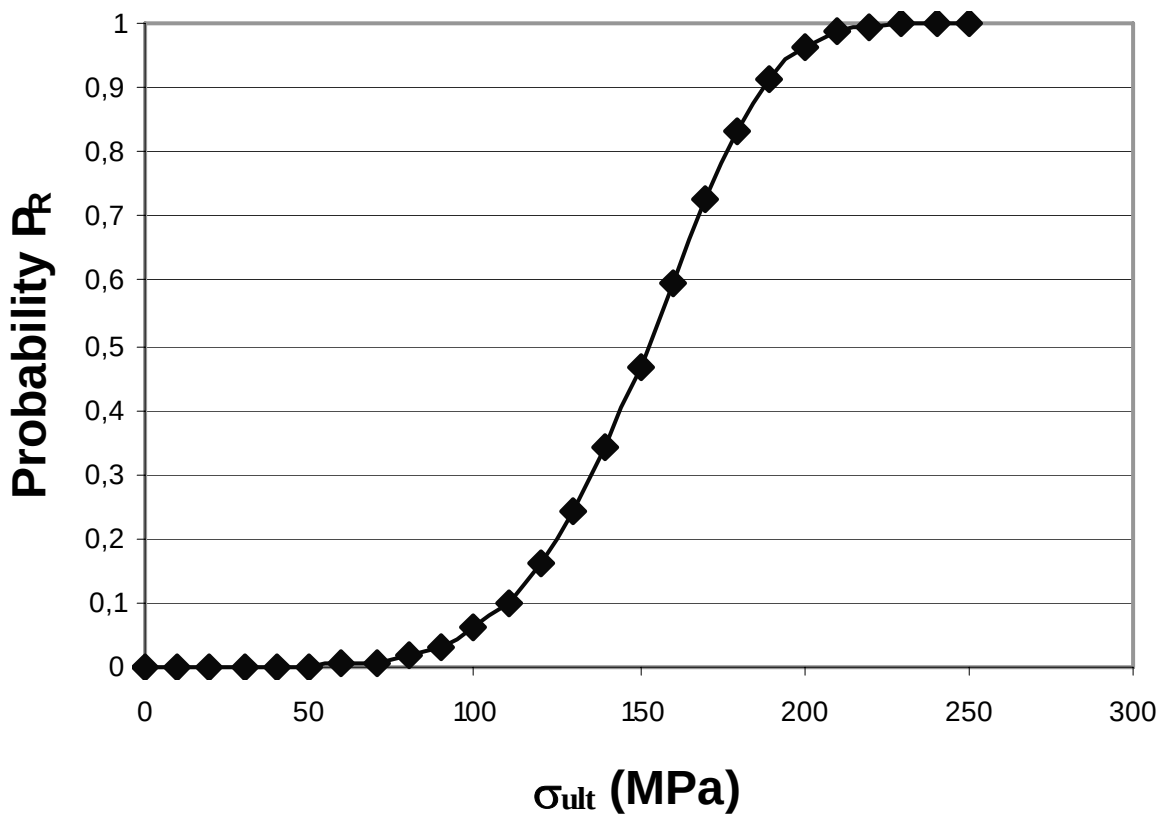

(a)

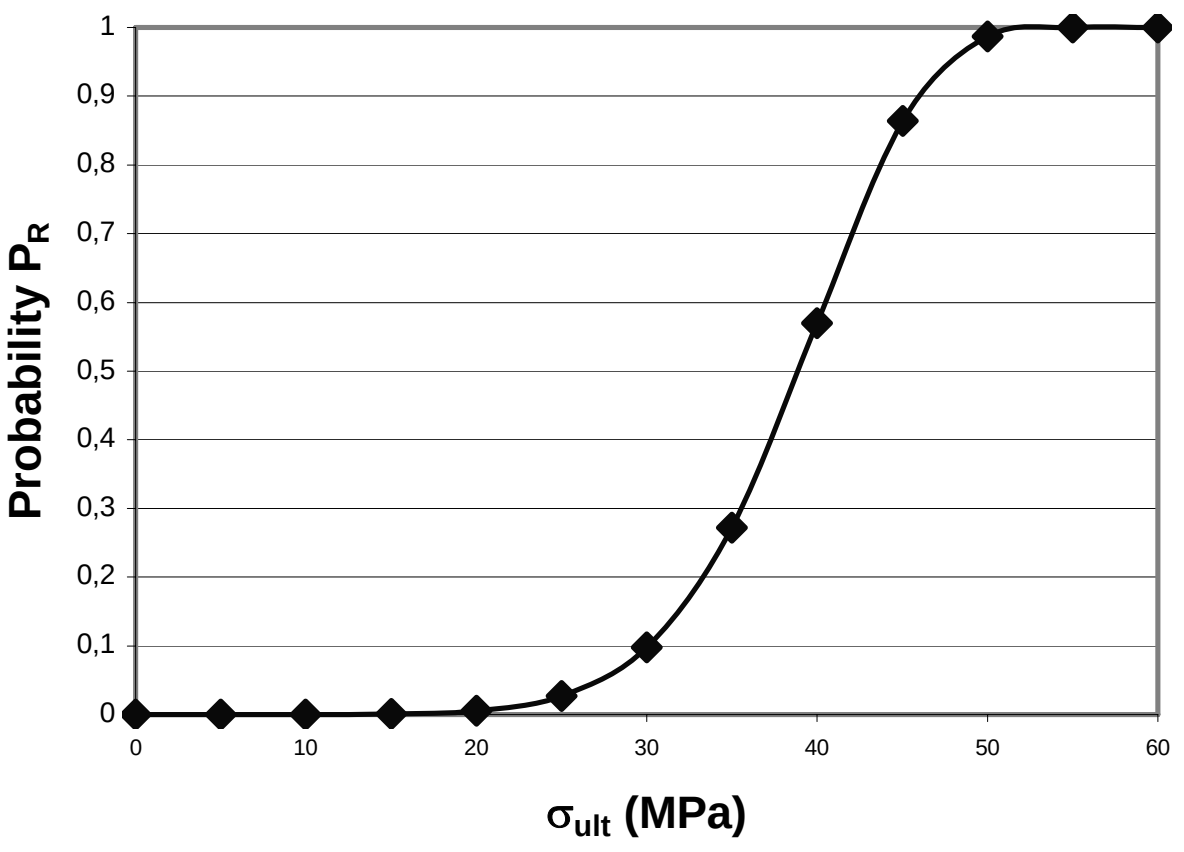

(b)

Figure 8

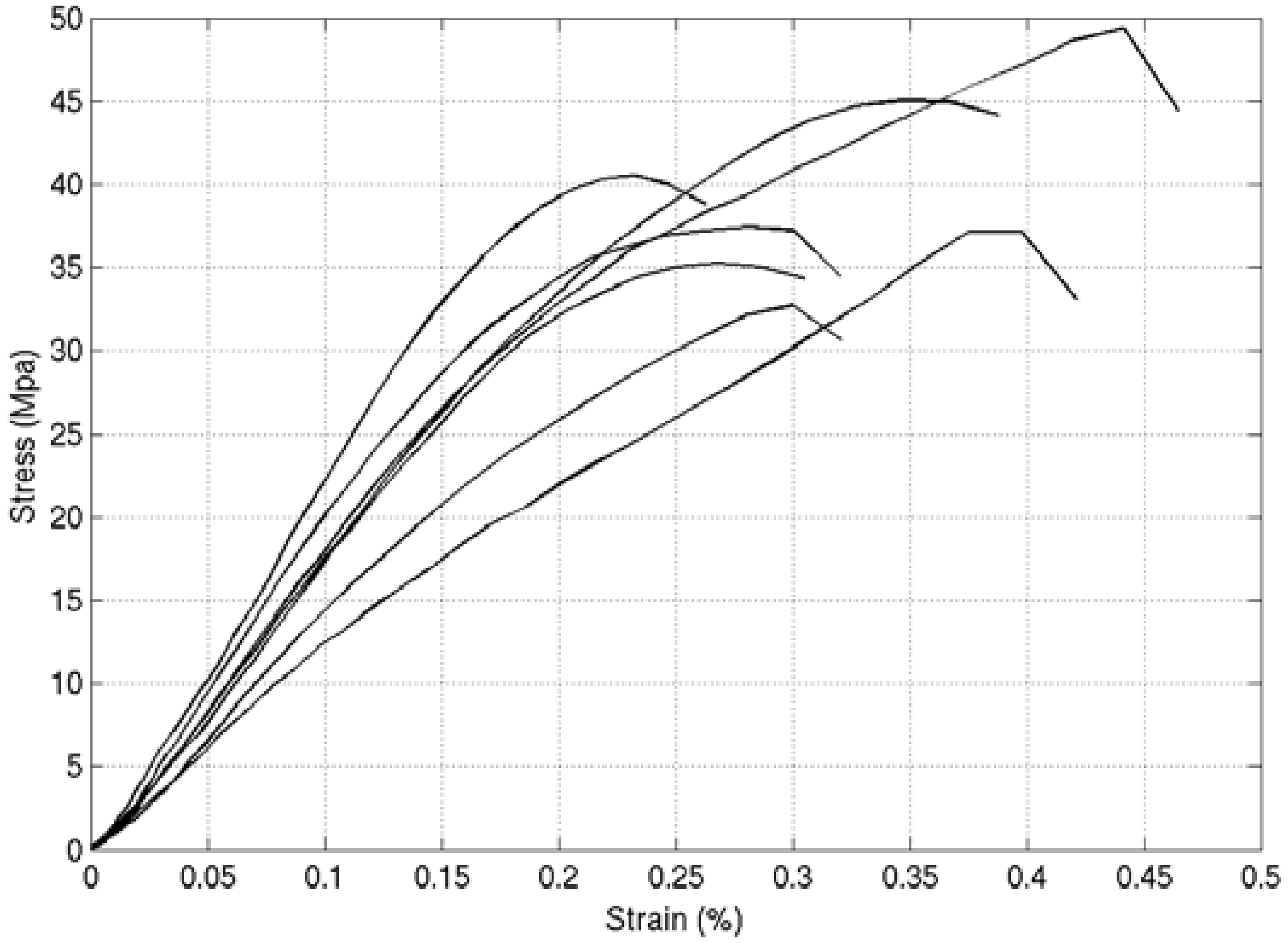

Figure 9

| | **Quasi static measurements** | **Measuring uncertainty of measurement** | **Dynamic measurements** | **Measuring uncertainty of measurement** |
|---|---|---|---|---|
| **E (GPa)** | $11.3 \leq E \leq 17.5$ | 2.7 % | $10 \leq E \leq 21.7$ | 6.6 % |
| **$\sigma_{rup}$(MPa)** | $105 \leq \sigma_{rup} \leq 193$ | 2 % | $33 \leq \sigma_{rup} \leq 50$ | 5.6 % |
| **$\varepsilon_{rup}$ (%)** | $0.93 \leq \varepsilon_{rup} \leq 1.8$ | 1 % | $0.23 \leq \varepsilon_{rup} \leq 0.44$ | 1 % |

Table 1